\documentclass[12pt]{iopart}
\usepackage{iopams}
\usepackage{epsfig}
\usepackage{amssymb}
\usepackage{amsfonts}
\usepackage{mathrsfs}
\usepackage{epsfig}
\usepackage{bm}

\begin{document}

\title{Influence of realistic parameters on state-of-the-art LWFA experiments}

\author{J. Vieira$^{1,5}$, S. F. Martins$^{1}$, F. Fi\'uza$^{1}$,C.K. Huang$^2$, W. B. Mori$^3$, S.P.D. Mangles$^4$, S. Kneip$^4$, S. Nagel$^4$, Z. Najmudin$^4$, L.O. Silva$^{1,6}$}

\address{$^1$~GoLP/Instituto de Plasmas e Fus\~{a}o Nuclear-Laborat\'orio Associado,  Instituto Superior T\'{e}cnico, 1049-001 Lisboa, Portugal}
\address{$^2$~Los Alamos National Laboratory, Los Alamos, New Mexico 87545}
\address{$^3$~Department of Physics and Astronomy, University of California, Los Angeles, CA 90095}
\address{$^4$~The Blackett Laboratory, Imperial College London, London, SW7 2AZ, United Kingdom}
\ead{$^5$jorge.vieira@ist.utl.pt, $^6$luis.silva@ist.utl.pt }

\begin{abstract}
We examine the influence of non-ideal plasma-density and non-Gaussian transverse laser-intensity profiles in the laser wakefield accelerator analytically and numerically. We find that the characteristic amplitude and scale length of longitudinal density fluctuations impacts on the final energies achieved by electron bunches. Conditions that minimize the role of the longitudinal plasma density fluctuations are found. The influence of higher order Laguerre-Gaussian laser pulses is also investigated. We find that higher order laser modes typically lead to lower energy gains. Certain combinations of higher order modes may, however, lead to higher electron energy gains.
\end{abstract}

\maketitle

\section{Introduction}
\label{sec:introduction}

Laser wakefield accelerators (LWFAs)~\cite{bib:tajima_prl_1979} have the potential to play an important role in a wide range of applications, including high-energy physics experiments (by providing more compact accelerators~\cite{bib:faure_nature_2004,bib:geddes_nature_2004,bib:mangles_nature_2004,bib:leemans_natphys_2006,bib:kneip_prl_2009}), nuclear physics experiments (by providing electron beams which can be used for the photonuclear activation of nuclei~\cite{bib:leemans_pop_2001,bib:reed_jap_2007}), and medical applications~\cite{bib:chiu_mp_2004} (by providing, for instance, compact x-rays sources~\cite{bib:kostyukov_pop_2003,bib:rousse_prl_2004}). For these applications, the recent developments of LWFAs, which currently deliver low divergence ($\sim 10~\mathrm{mrad}$), high-energy ($\sim 1~\mathrm{GeV}$), high charge ($\sim 10~\mathrm{pC}$) electron beams, with $\lesssim 10~\%$ energy spread, in externally-~\cite{bib:leemans_natphys_2006} or self-guided regimes~\cite{bib:kneip_prl_2009}, are placing strong emphasis on the shot-to-shot stability of the accelerators. It is therefore useful to understand the influence of shot-to-shot fluctuations on key outputs of LWFAs.

Shot-to-shot fluctuations of the plasma density and laser intensity profiles impact on the reproducibility of LWFAs. In particular, changes in the laser and plasma profiles may affect beam divergence, emittance, pointing stability, energy spread, and maximum energy. In this paper we will focus on their influence in energy gain. Previous research indicated that fluctuations in experimental plasma density profiles may prevent self-injection, thus constraining the applicability of LWFAs~\cite{bib:osterhoff_prl_2008}. However, plasma density ramps~\cite{bib:chien_prl_2005}, typically present in experiments with gas jets, can be used to control the phase-velocity of the plasma wave, and facilitate the self-injection of plasma electrons. Fluctuations of the laser intensity profile, which frequently exhibits higher-order modes, can affect the quality of the electron beams~\cite{bib:cummings_pop_2011}. Nevertheless, properly tailored higher-order laser modes can be used to lower the self-trapping thresholds~\cite{bib:michel_pop_2006} or increase the amplitude of the self-injected beam betatron oscillations~\cite{bib:glinec_epl_2008,bib:mangles_apl_2009}. Thus, either to obtain low shot-to-shot fluctuations on the outputs of LWFAs, or to obtain highly optimized acceleration regimes, an adequate control of the plasma density~\cite{bib:osterhoff_prl_2008}, and laser intensity~\cite{bib:mangles_pop_2007} is required. 


In order to investigate the influence of non standard plasma density and laser intensity profiles we use analytical modeling, and particle-in-cell simulations. Standard particle-in-cell algorithms are very computationally expensive. Reduced codes such as QuickPIC~\cite{bib:huang_jcp_2006}, or advanced PIC algorithms in boosted frames~\cite{bib:martins_natphys_2010,bib:martins_cpc_2010} are ideal to perform systematic parameter scans of the LWFA. In the boosted frame technique no physical approximations are involved. The associated computational gains result from bringing closer the disparate scales of LWFAs, i.e. the laser and the plasma wavelengths. The analysis of the relevant laser and plasma dynamics, however, is not straightforward, since  simultaneous events in the boosted frame occur at different instants of time in the laboratory frame~\cite{bib:martins_pop_2010}. The use of reduced models can circumvent these difficulties. The Quasi-Static Approximation (QSA), for instance, is widely used  to model LWFAs in reduced PIC codes~\cite{bib:sprangle_prl_1990,bib:mora_pop_1997}. It precludes the self-injection physics, but the dynamics of the laser, and acceleration of externally injected electron beams can be investigated systematically. Typically, the plasma response is determined through a Quasi-Static~\cite{bib:sprangle_prl_1990,bib:mora_pop_1997} plasma field solver, and the laser is advanced in the Ponderomotive Guiding Center Approximation (PGCA) ~\cite{bib:mora_pop_1997}. Simulations under the QSA and PGCA require lower spatial resolutions in comparison to standard full PIC codes. In addition, larger time steps can also be used~\cite{bib:fonseca_ppcf_2008, bib:fonseca_book}. Thus, reduced codes employing the QSA and PGCA can be much faster than standard full PIC codes. Here we employ the quasi-static, massively parallel, fully relativistic, PIC code QuickPIC which can be more than two orders of magnitude faster than standard full PIC codes~\cite{bib:huang_jcp_2006}.

In this paper, we investigate the influence of non-standard plasma and laser configurations in the outputs of LWFAs for a wide range of parameters relevant for state-of-the-art LWFAs. In Section~\ref{sec:standard}, a reference simulation using state-of-the-art laser and plasma parameters is presented and discussed. Then, in Sec.~\ref{sec:plasma}, the influence of inhomogeneous longitudinal plasma density perturbations is considered. It is shown that the impact of the density fluctuations can be minimized when the plasma-density variations are small in comparison to the background density, and when the typical wavelengths of the perturbations are much shorter than the acceleration distance. Section~\ref{sec:hermite} analyses the influence of non-Gaussian transverse laser intensity profiles in the acceleration. Analytical estimates for the energy gain are derived and confirmed with QuickPIC simulations. A parameter region where higher order modes can lead to higher energy gains was also identified. Finally the conclusions are stated.


\section{Self-guiding of 10 J laser pulses}
\label{sec:standard}

In this Section, the propagation of a 10 J laser pulse, with parameters close to Ref.~\cite{bib:kneip_prl_2009} is investigated using QuickPIC~\cite{bib:huang_jcp_2006}. 

In the baseline simulation conditions that we use in this paper, a laser pulse with central wavelength $\lambda_0=0.8~\mu\mathrm{m}$, normalized peak vector potential $a_0=3.85$, spot-size $W_0=19~\mu\mathrm{m}$ ($1/e$), and duration $\tau_0=55~\mathrm{fs}$ ($1/e$) is initialized at the entrance the pre-formed, transversely uniform plasma. The plasma density rises linearly to $n_0=5.7\times 10^{18}~\mathrm{cm}^{-3}$ for $0.065~\mathrm{cm}$. Uniform density is then used in the following $0.73~\mathrm{cm}$, falling linearly until the end of the plasma, during $0.12~\mathrm{cm}$. The simulation uses a window which moves at the speed of the light in vacuum (c), with $250~\mu\mathrm{m}\times250~\mu\mathrm{m}\times80~\mu\mathrm{m}$, divided into $512\times512\times256$ cells, with $4$ particles per cell. An electron beam with $Q=0.16~\mathrm{pC}$ is at the back of the first plasma wave, in a region where the accelerating fields are maximum. The charge of the beam is much lower than the beam loading charge~\cite{bib:tzoufras_prl_2008,bib:tzoufras_pop_2009}, in order to provide a quasi test particle acceleration regime. The electron beam density profile is given by:
\begin{equation}
\label{eq:beam}
N \propto \exp\left(-\frac{(\xi-\xi_0)^2}{2 \sigma_z^2}\right) \exp\left(-\frac{r^2}{2 \sigma_r^2}\right),
\end{equation}
where $\xi=z-c t$, $z$ is the propagation distance, $t$ the time, $\xi_0$ the center of the electron beam, $r=\sqrt{x^2+y^2}$ is the transverse coordinate, $\sigma_z=1.2~\mathrm{\mu m}$ is the length of the electron beam and $\sigma_r=1.2~\mathrm{\mu m}$ its width. The electron beam was initialized with a relativistic factor $\gamma=200$, with corresponding longitudinal velocity above the phase velocity of the plasma wave. Furthermore, the simulations included radiation damping, which results in a decelerating-like force, due to the betatron motion of the beam electrons betatron oscillations~\cite{bib:jackson}. For these parameters, we have found that radiation damping is negligible~\cite{bib:lu_prstab_2007}. 

\begin{figure}[t!hbp]
\begin{center}
\includegraphics[width=\columnwidth]{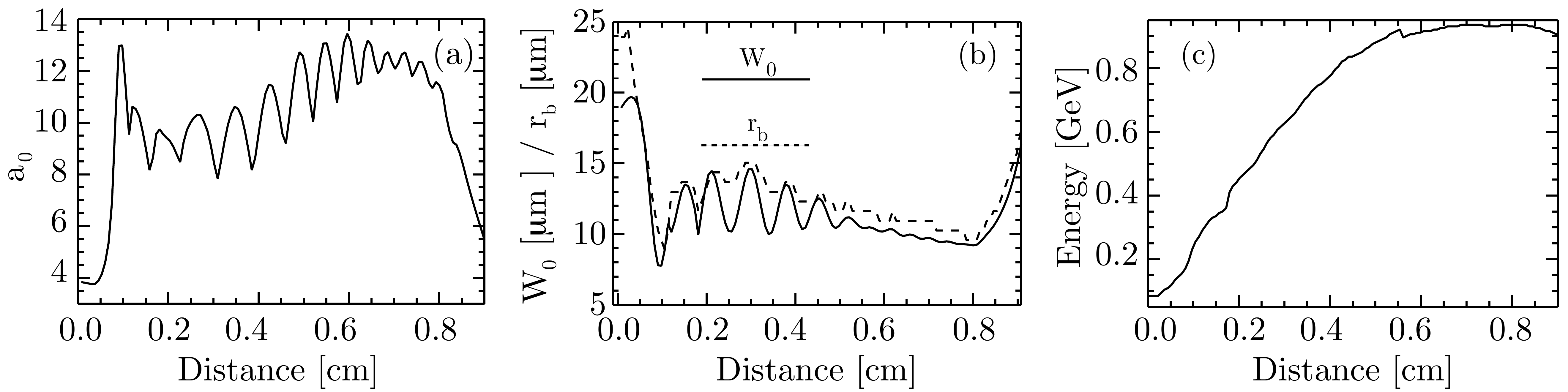}
\caption{\label{fig:standard} Baseline QuickPIC simulation results. (a) Evolution of the normalized peak laser pulse vector potential $a_0$ associated with an initially transversely Gaussian laser pulse. (b) Evolution of the laser pulse spot-size (solid line) and blowout radius (dashed line). The blowout radius corresponds to the location of the maximum plasma density for a fixed transverse slice, located roughly at the center of the laser pulse. (c) Evolution of the electron beam peak energy.}  
\end{center}
\end{figure}

The laser pulse peak $a_0$, spot-size $W_0$, blowout radius $r_b$, and electron beam peak energy, are depicted in Fig.~\ref{fig:standard}. For $z\leq 0.1~\mathrm{cm}$, strong laser spacial and temporal compression (Fig.~\ref{fig:standard}a) occurrs, where the $a_0$ increases from $a_0=3.85$ to $a_0= 13$~\cite{bib:tsung_pnas_2002}. Simultaneously, the $W_0$ decreases by the same factor, from $W_0=19~\mathrm{\mu m}$ to $W_0=7.5~\mathrm{\mu m}$ (Fig.~\ref{fig:standard}b). As a result, the ponderomotive force $F_p\propto a_0/W_0$ increases nearly by one order of magnitude, leading to complete electron blowout after $z=0.1~\mathrm{cm}$. This behavior, also present in \cite{bib:vieira_ieee_2008}, may be due to the mismatch between the initial laser spot-size and blowout radius $r_b$.

For $0.1~\mathrm{cm}<z<0.8~\mathrm{cm}$, both the $a_0$ and $W_0$ variations are smoother (Fig.~\ref{fig:standard}) indicating that matched conditions established self consistently. In average, $a_0$ increases from $a_0=8$ to $a_0=14$ until $z=0.8~\mathrm{cm}$, rapidly decreasing from then on. The increase of $a_0$ is accompanied by spot-size oscillations between $W_0=8~\mathrm{\mu m}$ and $W_0=16~\mathrm{\mu m}$, indicating that the body of the laser pulse is self-guided. Figures~\ref{fig:standard}a and \ref{fig:standard}b also show that, for $0.1~\mathrm{cm}<z <0.8~\mathrm{cm}$, both the $a_0$ and $W_0$ vary by more than $40\%$ in each oscillation. The corresponding variations of $r_b$, however, are lower than $10~\%$, thus guaranteeing a stable acceleration regime (in the blowout regime, the accelerating gradients are solely determined by $r_b$~\cite{bib:lu_prl_2006}). A closer investigation of Figs.~\ref{fig:standard}a and \ref{fig:standard}b reveals that $W_0$ and $a_0$ oscillate out of phase, i.e. $a_0$ is minimum for maximum $W_0$, and vice-versa, which lead to large variations of the ponderomotive force. However, the ponderomotive force at $r\simeq r_b$ is always similar, thus resulting in lower variations for $r_b$.

The stable accelerating fields lead to beam peak energies of almost 1 GeV at $z\sim0.7~\mathrm{cm}$ (in agreement with~\cite{bib:kneip_prl_2009}), which saturated from then on. The stabilization of the final electron beam peak energy was a result of the single beam electrons dynamics (not from the phase-space rotation due to the dephasing of the electron beam as in \cite{bib:tsung_prl_2004}): a fraction of the beam electrons are in regions of accelerating fields and gain energy. The remaining fraction are in a region of decelerating fields, and loose energy. As a result, the energy spectrum broadened with constant peak energy. 

These results establish the baseline conditions to analyze possible perturbations/deviations from the idealized parameters.

\section{Electron acceleration in inhomogeneous plasma density profiles}
\label{sec:plasma}

Nondeterministic plasma density fluctuations present in experiments impact on the acceleration gradients and final energies of electron bunches. Understanding the role of density non-uniformities in the LWFA is thus important to design stable accelerators. To this end we estimate analytically the single particle energy gain in a LWFA with small density perturbations. The expression for the energy gain of a trapped electron in the frame that moves with the laser group velocity $v_{g0}$, and in a reference plasma frequency $\omega_{p0}$, is given by:
\begin{equation}
\label{eq:energy_gain}
\Delta E = q \int_{-r_b}^{\xi_c} \frac{E_{\mathrm{accel}}(\xi_p)}{1-\beta_{\phi0}} \mathrm{d} \xi_p,
\end{equation} 
where $E_{\mathrm{accel}}$ is the accelerating field, $\xi_p = z_p - v_{g0}t = z_p - v_{\phi0} t \simeq z_p (1-\beta_{\phi0})$ is the distance of the externally injected electron to a point moving with the phase velocity of the plasma wave  $v_{\phi0}=\beta_{\phi0} c \equiv v_{g0}$, and $z_p=c t$ is the electron longitudinal position that moves at the speed of light. We consider that the electron, initially placed at $\xi_p=-r_b$, i.e. at the back of the bubble, accelerates in the positive $\xi_p$ direction until the center of the bubble $\xi_c$ is reached. At the center of the bubble, $E_{\mathrm{accel}}=0$, and the acceleration stops.

In order to determine an expression for $E_{\mathrm{accel}}$ we assume that the center of the bubble $\xi_c$ travels at the wake phase velocity in the blowout regime $v_{\phi}=c(1-\alpha \omega_{p}^2/\omega_0^2)$, where $\alpha = 3/2$, $\omega_0$ is the laser central frequency, $\omega_{p}=\omega_{p}(\xi_p)$ is the position dependent plasma wave frequency, and $\omega_0/\omega_p \gg 1$. Since in the blowout regime the accelerating fields are linear, we may write $E_{\mathrm{accel}}= (m_e \omega_p^2 / q) (\xi_p-\xi_c)/2$. In order to derive an expression for $\xi_c$ we assume that $\xi_c=0$ when $\xi_p = -r_b$. Hence the trajectory of the center of the bubble is $\xi_c = \int_{0}^{t_{\mathrm{accel}}}(v_{\phi}-v_{\phi0})\mathrm{d}t$, where $t_{\mathrm{accel}}$ is the acceleration time. Since a trapped electron moves at $c$, $\xi_p=(c-v_{\phi0}) t$, and hence $\xi_c(\xi_p)=\int_{-r_b}^{\xi_p} \left[\omega_{p}(\xi)^2/\omega_{p0}^2 -1\right] d\xi$. In an ideal scenario where the plasma is uniform ($\omega_p(\xi_p)$ is constant) the acceleration stops at $\xi_c=0$. In general, if $\omega_p(\xi_p)$ is not constant $\xi_c$ may be different than zero. As a result, Eq.~(\ref{eq:energy_gain}) can be re-written as:
\begin{equation}
\label{eq:energy_gain_i}
\Delta E = - \frac{3}{2} m_e \int_{-r_b}^{\xi_c} \frac{\omega_0^2 \omega_p^2(\xi_p) \left[\xi_p-\xi_c(\xi_p)\right]}{2 \omega_{p0}^2} \mathrm{d} \xi_p,
\end{equation} 

To further evaluate Eq.~(\ref{eq:energy_gain_i}) we assume small plasma density perturbations such that $\omega_p^2(\xi_p) = \omega_{p0}^2 [1+ \delta\omega_p(\xi_p)]$ where $\delta \omega_p(z_p)\ll 1$ is a small perturbation. For the sake of simplicity we also assume that the plasma density varies sinusoidally, and that the average plasma density (frequency) is $n_0$ ($\omega_{p0}$). Hence, $\delta\omega_p(\xi_p) = \delta \cos(k_1 \xi_p + \phi)$ where $\delta\ll 1$ is the amplitude of the sinusoidal plasma density perturbation. The initial phase $\phi=-\pi/2 - k_1 \xi_{p0}$ guarantees that $\delta \omega_p(-r_b) = 0$, and that the plasma density rises for $\xi_p \gtrsim -r_b$ if $\delta > 0$ ($\xi_{p0}=-r_b$ is the initial electron bunch position). The wavenumber $k_1$ is the typical wave number of the non-uniformity in the frame that travels with $v_{\phi0}$. In the laboratory frame coordinates ($z_p(\xi_p),t$), $k_1 =(2/3) k_1^{\mathrm{lab}} \omega_0^2/\omega_{p0}^2$.

\begin{figure}[htbp]
\begin{center}
\includegraphics[width=0.6 \columnwidth]{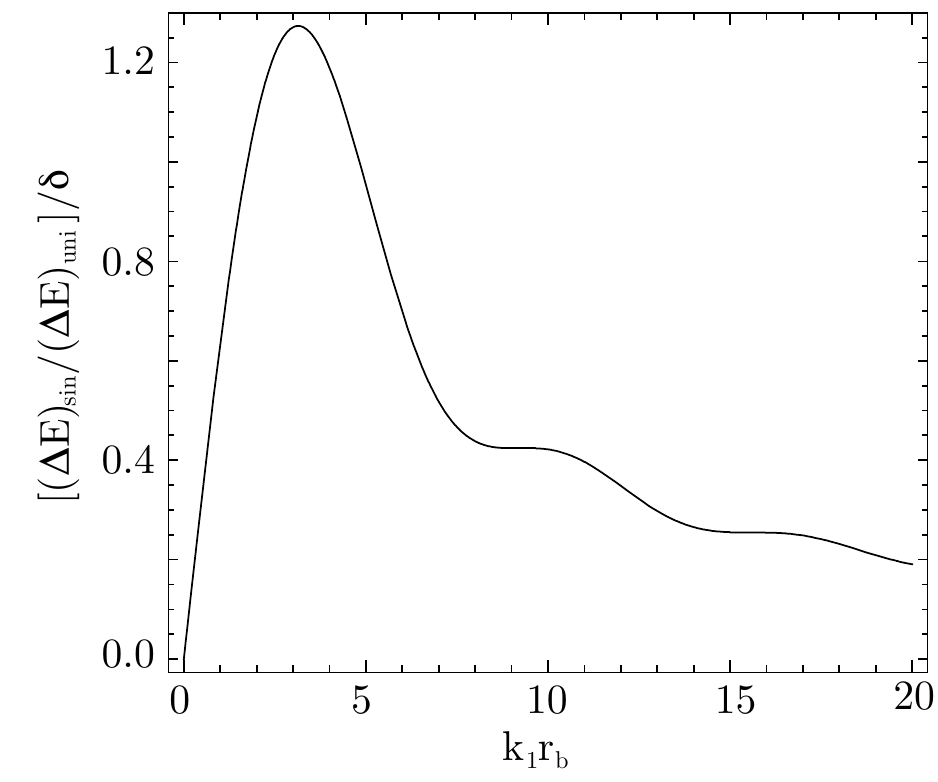}
\caption{\label{fig:energy_var} Normalized ratio $[(\Delta  E)_{\mathrm{sin}}/(\Delta  E)_{\mathrm{uni}}]/\delta$ as a function of $k_1 r_b$. Maximum $[(\Delta  E)_{\mathrm{sin}}/(\Delta  E)_{\mathrm{uni}}]/\delta$ occurs for $r_b k_1=\pi$, where $[(\Delta  E)_{\mathrm{sin}}/(\Delta  E)_{\mathrm{uni}}]/\delta = 4/\pi \simeq 1.27$.}
\end{center}
\end{figure}

For the sake of simplicity, we consider that the upper limit of the integral in Eq.~(\ref{eq:energy_gain_i}) is $\xi_c=0$. Retaining the leading order terms on the order of $\mathcal{O}(\delta)$ leads to: 
\begin{equation}
\label{eq:energy_gain_ii}
\Delta E = m_e c^2 \left[\frac{3}{8}\frac{\omega_0^2}{\omega_{p0}^2}\frac{r_b^2 \omega_{p0}^2}{c^2} - \frac{3 \delta}{2} \frac{k_{p0}}{k_1}\frac{\omega_0^2}{\omega_{p0}^2}\left(\frac{r_b\omega_{p0}}{c} - \frac{k_{p0}}{k_1} \sin\left(r_b k_1\right)\right)\right],
\end{equation}
The first term on the right-hand-side of Eq.~(\ref{eq:energy_gain_ii}), $(\Delta E)_{\mathrm{uni}} = m_e c^2 (3/8) (r_b\omega_p/c)^2 \omega_0^2/\omega_{p0}^2$, corresponds to the maximum energy gain in an uniform plasma. Making the substitution $r_b =2 \sqrt{a_0} c/\omega_p$ yields $(\Delta E)_{\mathrm{uni}} = m_e c^2 (3/2) a_0 \omega_0^2/\omega_{p0}^2$ which is close to the result from Ref.~\cite{bib:lu_prstab_2007}. The differences in the numerical factors are due to the fact that Ref.~\cite{bib:lu_prstab_2007} considers a constant average $E_{\mathrm{accel}}$. Instead, Eq.~(\ref{eq:energy_gain_ii}) uses linear accelerating fields. The second term, $(\Delta E)_{\mathrm{sin}} = - m_e c^2(3 \delta \omega_0^2 k_{p0}/2 \omega_{p0}^2 k_1) \left[r_b \omega_{p0}/c - (k_{p0}/k_1)\sin\left(r_b k_1\right)\right]$ are corrections associated with the density non-uniformities. These corrections are due equally to the variations on the accelerating gradients, and to the variations of the laser group velocity through fluctuations of $\xi_c$. According to Eq.~(\ref{eq:energy_gain_ii}), the ratio $(\Delta  E)_{\mathrm{sin}}/(\Delta  E)_{\mathrm{uni}}$ is:
\begin{equation}
\label{eq:ratio}
\frac{(\Delta  E)_{\mathrm{sin}}}{(\Delta  E)_{\mathrm{uni}}} = \frac{4 \delta [r_b k_1 -\sin(r_b k_1)]}{r_b^2 k_1^2}.
\end{equation}
Equation~(\ref{eq:ratio}) is plotted in Fig.~\ref{fig:energy_var} as a function of $r_b k_1$. The ratio $(\Delta E)_{\mathrm{sin}}/(\Delta E)_{\mathrm{uni}}$ is directly proportional to $\delta$. Larger final energy fluctuations are hence expected in plasmas with larger density fluctuations. In addition, the sign of $\delta$ determines the sign of $(\Delta E)_{\mathrm{sin}}$. A density increase (reduction) at the start of the acceleration may then lead to larger (lower) energy gains in comparison to uniform plasma densities. This is associated with the fact that the acceleration gradients decrease as the acceleration progresses. Hence, the density fluctuations at the start of the acceleration will be more important than those occurring later. The initial density fluctuations then give the dominant contribution for $(\Delta E)_{\mathrm{sin}}$.

Equation~(\ref{eq:ratio}) also shows that $(\Delta E)_{\mathrm{sin}}$ depends on the typical wavenumber of the density non-uniformities. For $r_b k_1 \ll \pi$ the density fluctuations can be neglected because their amplitude is arbitrary small in this limit. For $r_b k_1 = \pi$, $(\Delta E)_{\mathrm{sin}}/(\Delta E)_{\mathrm{uni}} = 4 \delta/\pi\simeq 1.27\delta$ is maximum. In the laboratory frame coordinates ($z,t$) this corresponds to a wavenumber $k_1^{\mathrm{lab}}=(3 \pi/2) \omega_{p0}^2/(r_b \omega_0^2)$, and to a wavelength $\lambda_1^{\mathrm{lab}} = 2 \pi / k_1^{\mathrm{lab}} = (4/3) r_b \omega_0^2 / \omega_p^2$. The total acceleration length in the matched blowout regime is $L_{\mathrm{accel}} \simeq (2/3)\omega_0^2 r_b/\omega_{p0}^2$~\cite{bib:lu_prstab_2007}. Thus $\lambda_1^{\mathrm{lab}} \simeq 2 L_{\mathrm{accel}}$. In the limit $r_b k_1 \gg \pi$, or equivalently $\lambda_1^{\mathrm{lab}} \ll 2 L_{\mathrm{accel}}$, $(\Delta E)_{\mathrm{sin}}\rightarrow 0$ because the effects associated with the density non-uniformities in the acceleration will average to zero. 

\begin{figure}[htbp]
\begin{center}
\includegraphics[width=\columnwidth]{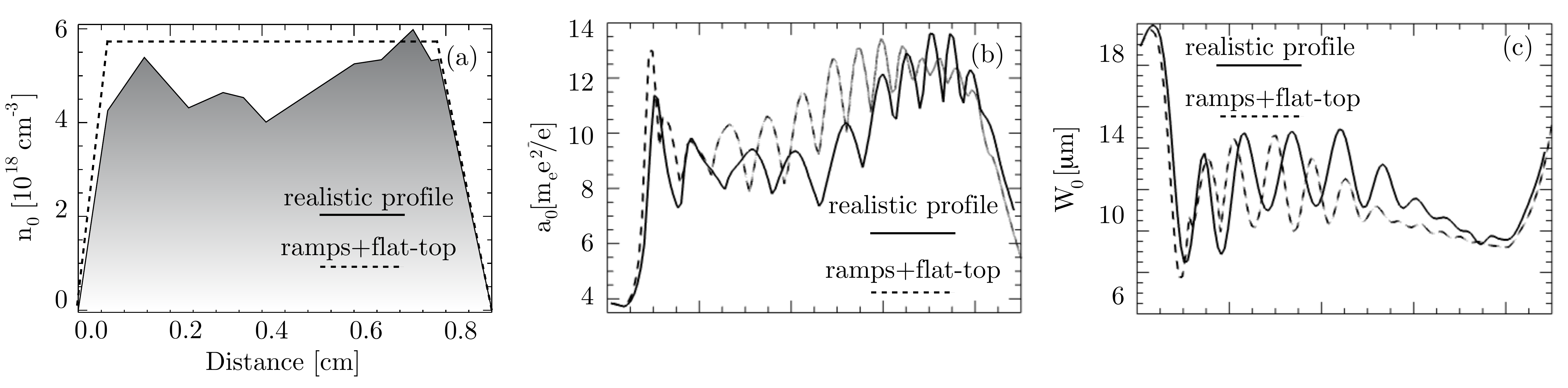}
\caption{\label{fig:non_ideal_density}QuickPIC simulation results comparison between an experimental (solid-black line) and idealized (dash-black line) plasma density profiles (a) Realistic density profile, with small ($\lesssim 30\%$) fluctuations from the idealized profile, characterized by two ramps at the edges of a uniform plasma density region. (b) comparison between the peak normalized vector potentials, and (c) comparison of the evolution of the laser pulse spot-size.}  
\end{center}
\end{figure}

Equation~(\ref{eq:energy_gain_ii}) assumes that the density variations are sinusoidal, that the wakefield amplitude is constant, and that the acceleration distance does not change, and neglects pump-depletion and beam loading effects. In order to verify that Eq.~(\ref{eq:energy_gain_ii}) holds even when these approximations are not verified, we performed particle-in-cell simulations in QuickPIC in conditions relevant for experiments. The baseline simulation parameters for the laser and plasma are identical to those of Sec.~\ref{sec:standard}. We first investigate the laser dynamics in the experimental density profile shown in Fig.~\ref{fig:non_ideal_density}a, which was obtained from interferometric measurements of the plasma channel profile during the experiment reported in \cite{bib:kneip_prl_2009}. Figures~\ref{fig:non_ideal_density}b, and \ref{fig:non_ideal_density}c show that the global evolution of $a_0$ and $W_0$ in the inhomogeneous density profile is identical to that described in Sec.~\ref{sec:standard}: the propagation is dominated by strong initial self-focusing that leads to the complete blowout of plasma electrons, followed by a self-guided propagation stage. 

In spite of the similar laser dynamics associated with the profiles represented in Fig.~\ref{fig:non_ideal_density}a, two differences can still be mentioned: first of all, the variation of the laser $a_0$ ($W_0$) in Sec.~\ref{sec:standard} is systematically higher (lower) than the corresponding variations associated with the experimental profile of Fig.~\ref{fig:non_ideal_density}a. Moreover, the frequency of the $a_0$ ($W_0$) oscillations is also higher in Sec.~\ref{sec:standard} than by using the experimental profile. These effects are both due to the higher plasma density used in Sec.~\ref{sec:standard}. On one hand, the use of higher densities leads to stronger laser pulse ponderomotive forces, and self-focusing, and thus to higher peak  $a_0$'s. Consequently, the frequencies of oscillation of both $W_0$ and $a_0$ are higher. Furthermore, the frequency shifts are stronger and faster for high densities, which further increases the peak $a_0$'s obtained using the experimental profile from Fig.~\ref{fig:non_ideal_density}.

\begin{figure}[htbp]
\begin{center}
\includegraphics[width=\columnwidth]{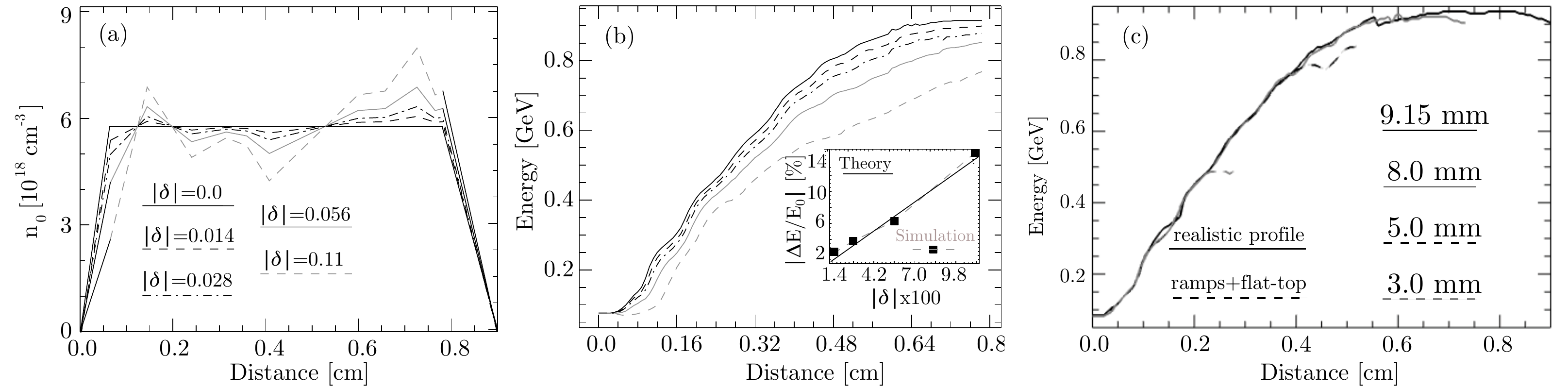}
\caption{\label{fig:ene_peaks} Energy gain by externally injected electrons propagating in non-uniform plasma profiles with different amplitudes of the density variations and plasma lengths. (a) Simulated longitudinal plasma density profiles as a function of the propagation distance. (b) Corresponding energy gain by externally injected electrons. The inset in (b) shows a comparison between the simulated energy gain variation after 0.8 cm of propagation with the theoretical estimates. (c) Energy gain by externally injected electrons using different plasma lengths with ramps identical to those shown in (a).}
\end{center}
\end{figure}

Simulation results from a systematic parameter scan with different amplitudes of the plasma density non-uniformities is shown in Fig.~\ref{fig:ene_peaks}. The simulated density profiles are variations of that of Fig.~\ref{fig:non_ideal_density}a with different $|\delta|$'s. The value of $\delta$ corresponds to the standard deviation of the density variations along the plasma multiplied by $\sqrt{2}$ such that it is consistent with the amplitude of the sine function used in the analytical model. In addition, the average plasma density is identical for each simulated profile, and the test electron bunches were injected at the same distance from the laser in each simulation. 

Figure~\ref{fig:ene_peaks}b shows the energy gain of external electron bunches as a function of the propagation distance for each of the profiles of Fig.~\ref{fig:ene_peaks}a. Figure~\ref{fig:ene_peaks}b confirms that larger plasma density fluctuations lead to larger energy gain variations in comparison to uniform plasma densities. In addition, larger $|\delta|$'s lead to lower energy gains since the first part of the propagation occurs in regions where the plasma density is lower. These findings agree qualitatively with Eq.~(\ref{eq:energy_gain_ii}). In order to compare the model and simulations quantitatively we plot in the inset of Fig.~\ref{fig:ene_peaks}b the theoretical value for the maximum $(\Delta E)_{\mathrm{sin}}/(\Delta E)_{\mathrm{uni}}=4 \delta/\pi$, and the maximum energy gain for each simulation as a function of $\delta$. The simulation results follow the theoretical results for maximum $(\Delta E)_{\mathrm{sin}}/(\Delta E)_{\mathrm{uni}}$.

We also analyzed the role of the plasma density ramps in the acceleration gradients. Simulations results with ramp lengths identical to those of Figs.~\ref{fig:non_ideal_density}a and \ref{fig:ene_peaks}a but with different plasma lengths are illustrated in Fig.~\ref{fig:ene_peaks}c. These simulations indicate that the energy gain does not depend on the presence of the plasma ramps. Naturally, larger plasma ramps will affect final energy gains. However, our results illustrate that in the typical experimental considered conditions, electron energy gains are insensitive to the presence of the plasma ramps.

All simulations showed large energy spreads on the order of 100\% at the end of the plasma. This is due to the fact that initially the wake is not beam loaded. We use low beam charges, such that the acceleration occurs in a test-particle-regime. Consequently the acceleration differed for particles at different longitudinal positions. The results from these simulations then did not show a clear influence of $\delta$ on the particle energy spread. However, it is expected that plasma density non-uniformities will influence the energy spread when the plasma wave is beam loaded.

\section{Role of higher-order Laguerre-Gaussian laser pulses in the LWFA}
\label{sec:hermite}

The presence of higher order modes in the transverse laser intensity profile is ubiquitous in LWFAs. Hence, understanding the role of higher order modes in the laser propagation and in the final electron bunch energy is important to plan and interpret experiments. In this section we will examine LWFAs excited by radially symmetric laser pulses with higher order Laguerre-Gaussian modes~\cite{bib:tzeng_phd}. In vacuum, the slowly varying envelope of the normalized vector potential associated with an arbitrary laser beam profile can be written as a sum of Laguerre-Gaussian modes as:
\begin{eqnarray}
\label{eq:laguerre}
\mathbf{a} & = & f(z)\sum_{l,p} \mathbf{a}^{(l,p)} \left(\frac{r \sqrt{2}}{W_0(z)}\right)^{|l|}\exp\left(-\frac{r^2}{W_0^2(z)}\right) \mathrm{L}_p^{|l|}\left(\frac{2 r^2}{W_0^2(z)}\right) \times \nonumber \\
&\times &\exp\left(i \frac{k_0 r^2}{2 R(z)}\right) \exp\left(i l \phi\right),
\end{eqnarray}
where $f(z)$ is the longitudinal laser profile, $\phi$ is the azimuthal coordinate, $W_0(z)=W_0 \sqrt{1+z^2/z_r^2}$ is the vacuum laser spot size as a function $z$, $z_r = k_0 W_0^2/2$ is the Rayleigh length, $R(z) = z\left(1+z_r^2/z^2\right)$ is the radius of curvature of the laser wavefronts, and $L_p^{|l|}$ are generalized Laguerre polynomials of degree $p$. Since we analyze the propagation of radially symmetric laser beams we consider $l=0$. Eq.~(\ref{eq:laguerre}) then becomes:
\begin{equation}
\label{eq:laguerre_symmetric}
\mathbf{a}=f(z)\sum_{p} \mathbf{a}^{(p)} \exp\left(-\frac{r^2}{W_0^2(z)}\right) \mathrm{L}_p\left(\frac{2 r^2}{W_0^2(z)}\right) \exp\left(i \frac{k_0 r^2}{2 R(z)}\right),
\end{equation}
where $\mathrm{L}_p\equiv \mathrm{L}_p^{(0)}$. Equation~(\ref{eq:laguerre_symmetric}) shows how the different combinations of higher order Laguerre-Gaussian modes lead to different radial laser intensity, energy and power distributions. Since the laser power $P$ is conserved for each transverse laser slice, we assume that initial laser power distribution determines the electron energy gain. The energy gain $\Delta E$ in the blowout regime as a function $P$ is~\cite{bib:lu_prstab_2007}:
\begin{equation}
\label{eq:energy}
\Delta E [GeV] = 1.7 \left(\frac{P[\mathrm{TW}]}{100}\right)^{1/3}\left(\frac{10^{18}}{n_0[\mathrm{cm}^{-3}]}\right)^{2/3}\left(\frac{0.8}{\lambda_0 [\mu\mathrm{m}]}\right)^{4/3},
\end{equation}
where $\lambda_0 = 2 \pi c/k_0$ is the central laser wavelength.

The fraction of the laser power which is far from the blowout region may diffract, not contributing to wake excitation and particle acceleration. Thus, we consider that the only laser power useful for particle acceleration is that contained within $r\lesssim r_b$. According to Fig.~\ref{fig:standard} this corresponds to the laser power contained within $r\lesssim W_0$. The power at the focal plane contained within a laser spot size for lasers with $p=0$ (Gaussian beam) and $p=1$ is:
\begin{eqnarray}
\label{eq:power}
P & \propto &  \int_{0}^{W_0}  r \left[a^{(0)}+ a^{(1)} \exp\left(-\frac{r^2}{W_0^2}\right) \mathrm{L}_p\left(\frac{2 r^2}{W_0^2}\right)\right]^2 \mathrm{d}r \nonumber \\
& \propto & \frac{W_0^2 \left(a^{(0)2}+a^{(1)2}\right)}{4} -\frac{W_0^2 \left(a^{(0)2} - 4 a^{(0)} a^{(1)} + 5 a^{(1)2}\right)}{4 \e^2},
\end{eqnarray}
To make quantitative predictions we assume that the laser pulse energy $E^{\mathrm{laser}}$ is constant for any combination of $a^{(0)}$ and $a^{(1)}$. According to Eq.~(\ref{eq:laguerre_symmetric}), the laser energy is proportional to $E^{\mathrm{laser}}\propto (a^{(0)2}+a^{(1)2})\equiv a_0^2$, where $a_0$ is the normalized vector potential of a purely Gaussian laser beam with $W_0$ and $f(z)$ identical to the higher order mode. Thus, $a^{(0)}=\sqrt{a_0^2-a^{(1)2}}$. Making use of the latter when inserting Eq.~({\ref{eq:power}}) into Eq.~(\ref{eq:energy}), we can calculate the ratio $\Delta E^{(1)}/\Delta E^{(0)}$ given by
\begin{equation}
\label{eq:energy_gain_laguerre}
\frac{\Delta E^{(1)}}{\Delta E^{(0)}} = \left(1-\frac{4 a^{(1)2}-4 a^{(1)} \sqrt{a_0^2-a^{(1)2}}}{a_0^2 \left[\e^2-1\right]}\right)^{1/3},
\end{equation}
where $\Delta E^{(1)}$ ($\Delta E^{(0)}$) is the energy gain of the laser pulse with higher order (purely Gaussian) modes. Note that when the higher order laser pulse power inside $r<W_0$ is the same as a Gaussian pulse then $\Delta E^{(1)}/\Delta E^{(0)}=1$. Equation~(\ref{eq:energy_gain_laguerre}) can be used to examine the impact of the first $p=0$ and $p=1$ Laguerre-Gaussian modes in the acceleration of electrons, and can be readily generalized including the presence of modes with $p>1$. It shows that the presence of higher order Laguerre-Gaussian modes generally leads to lower energy gains in comparison to Gaussian laser pulses. This is because the laser power is typically distributed around wider radial regions for lasers with $a^{(1)}\ne 0$. Minimum energy gains occur when $a^{(1)}= -(a_0/2)\sqrt{2+\sqrt{2}}\simeq 0.92 a_0$ for which $\Delta E^{(1)}/\Delta E^{(0)}=[(\e^2-3-2\sqrt{2})/(\e^2-1)]^{1/3}\approx 0.625$. Interestingly, Eq.~(\ref{eq:power}) also predicts that the energy gain in the range $0<a^{(1)}<a_0/\sqrt{2}$ is higher than that of a Gaussian pulse. Maximum energy gain $\Delta E^{(1)}/\Delta E^{(0)}$ is obtained when $a^{(1)}= (a_0/2)\sqrt{2-\sqrt{2}}\simeq 0.38 a_0$ for which $\Delta E^{(1)}/\Delta E^{(0)}=[(\e^2+2\sqrt{2}-3)/(\e^2-1)]^{1/3}\approx 1.04$. In this case more laser power is contained within $r<W_0$ in comparison to a purely Gaussian laser leading to $\Delta E^{(1)}/\Delta E^{(0)}>1$. 
 
\begin{figure}[t!hbp]
\begin{center}
\includegraphics[width=\columnwidth]{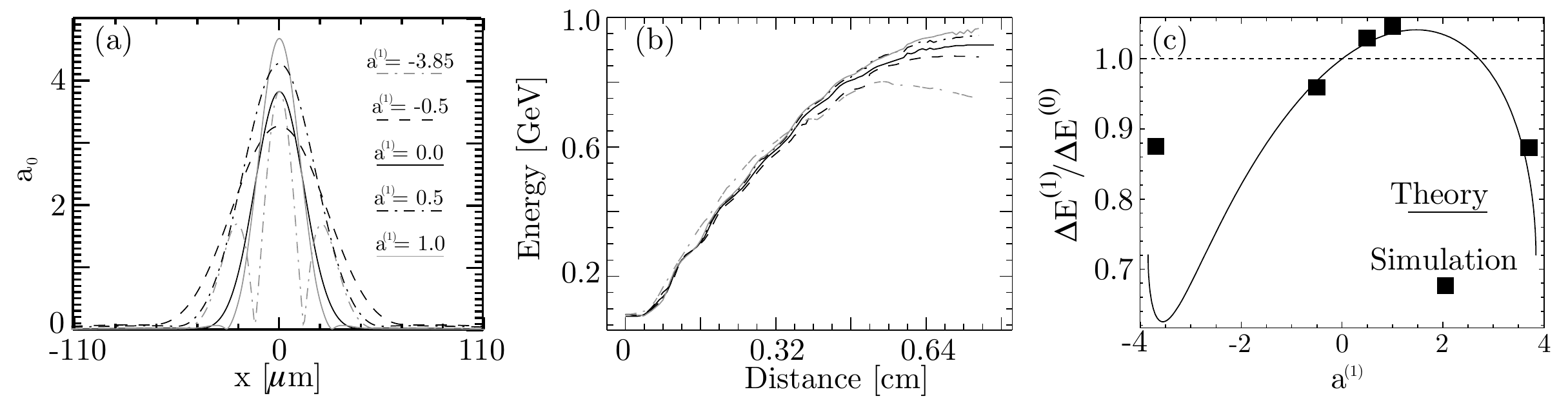}
\caption{\label{fig:laguerre} Influence of higher order Laguerre-Gaussian laser beam modes in the energy gain in the LWFA. (a) Initial central laser lineouts for each simulated scenario. (b) Corresponding energy gain as a function of the propagation distance. (c) Comparison between the analytical model Eq.~(\ref{eq:energy_gain_laguerre}) and the simulation results.}  
\end{center}
\end{figure}

In order to confirm these findings we performed two sets of QuickPIC simulations using lasers with higher order Laguerre-Gaussian modes. The baseline parameters (laser energy, spot-size, duration, and plasma density) are similar to those presented in Sec.~\ref{sec:standard}. In the first set of simulations we analyzed electron energy gain by lasers with $p=0$ and $p=1$. This permitted to make direct comparisons with Eq.~(\ref{eq:energy_gain_laguerre}). Results are summarized in Fig.~(\ref{fig:laguerre}). Very good agreement was found between the analytical model (cf. Eq.~(\ref{eq:energy_gain_laguerre})) and simulation results. In particular, simulations confirmed that higher order Laguerre-Gaussian laser beams can lead to higher electron energies. Simulations also revealed that the theory underestimates the energy gain for $(a^{(0)},a^{(1)})=(0,\pm 3.85)$. Simulations in this scenario show that a fraction of the laser initially at $r\gtrsim W_0$ self-focuses and enters the blowout region during the propagation. This mechanism increases the total laser power inside $r<W_0$ that can be used in particle acceleration, but is not accounted for in Eq.~(\ref{eq:energy_gain_laguerre})

In a second set of QuickPIC simulations we test the model using ``realistic'' laser pulse profiles. The simulated laser parameters are a numerical fit of measurements of the focal spot laser intensity distribution from~\cite{bib:kneip_prl_2009}. The simulation uses $a^{(0)}=1.9$, $a^{(1)}=-0.44$, $a^{(2)}=1.14$, $a^{(3)}=0.05$, $a^{(4)}=0.46$, $a^{(5)}=0.13$, $a^{(6)}=0.33$, $a^{(0)}=-0.39$, and $W_0=30~\mu\mathrm{m}$. The plasma density, laser duration and laser energy are similar to the baseline simulation parameters of Sec.~\ref{sec:standard}.

\begin{figure}[htbp]
\begin{center}
\includegraphics[width=\columnwidth]{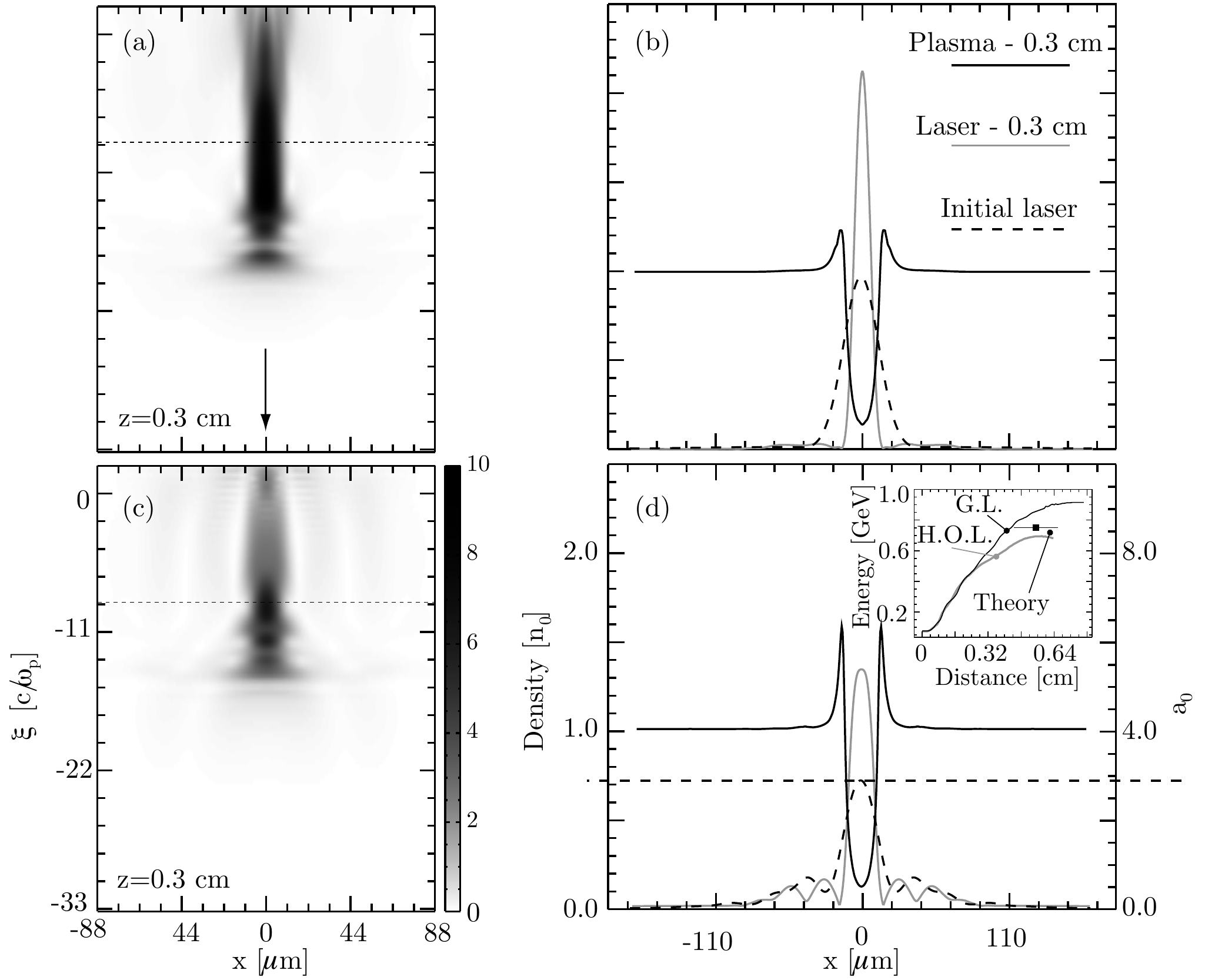}
\caption{\label{fig:laguerre_experimental} QuickPIC simulation results comparing a scenario using a purely Gaussian laser beam (a)-(b), and a laser with higher order Laguerre-Gaussian modes (c)-(d). (a) is a 2D slice at y=0 illustrating the laser vector potential after 0.3 cm. The arrow indicates the laser propagation direction. The dashed line indicates the position where a transverse line-out was taken in (b). (b) Lineout of the plasma density, and laser vector potential after 0.3 cm. The dashed curve is the initial laser profile. (c) is a 2D slice at y=0 illustrating the profile of a fit to an experimental laser vector potential after 0.3 cm. (d) lineouts of the plasma and laser profiles. The laser profile (dashed-line) is characterized by a central filament, with several filaments in the side. The inset in (d) shows the energy gain of externally injected electrons as a function of the propagation distance. The theoretical prediction is shown by the dark square, the gray curve represents the simulation results with higher order laser modes (H.O.L.), and the black curve represents the simulation results with a purely Gaussian Laser (G.L.).}  
\end{center}
\end{figure}

Figure~\ref{fig:laguerre_experimental} shows the main simulation results. For comparison, Fig.~\ref{fig:laguerre_experimental}a plots the Gaussian laser profile associated with the baseline simulation parameters. The corresponding plasma density profile is also illustrated in Fig.~\ref{fig:laguerre_experimental}b. These figures reveal that higher order modes form during the propagation even if the initial laser profile is Gaussian. This is because higher order modes appear naturally when Gaussian lasers propagate in non-parabolic plasma density profiles such as those associated with the blowout regime.

Figure~\ref{fig:laguerre_experimental}c-d show simulation results using the fit to the experimental laser profile including higher order modes. Since the initial laser profile already contains higher order Laguerre-Gaussian modes, the laser filaments at $z=0.3~\mathrm{cm}$ are much more pronounced than in Fig.~\ref{fig:laguerre_experimental}a-b. As the laser propagates it self-focuses leading to the generation of non-linear plasma waves in the blowout regime. A sharp electron interface that separates the bubble from the surrounding plasma is then created. This thin plasma sheet at $r=r_b$ bends the laser pulse wavefronts outwards for $r\gtrsim W_0\sim r_b$, and inwards for $r\lesssim W_0 \sim r_b$, which thus confines the central region of the laser pulse inside the blowout region, and prevents the entrance of the side filaments inside it. The blowout region thus acts as a spatial filter of the laser pulse. Most of the laser light outside the blowout region is then lost via diffraction and pump depletion. This process confirms that only a fraction of the initial laser pulse power or energy is used to create the blowout and accelerate electrons. As a result, the final electron beam energies are lower in comparison with the results from Sec.~\ref{sec:standard}. The electron energy gain as a function of the propagation distance is shown in the inset of Fig.~\ref{fig:laguerre_experimental}d. Despite the theory does not account for the details of the wake excitation by lasers with higher order modes, and for the laser evolution, it agrees very well with simulations.

The final energy achieved by LWFAs driven by lasers with higher order modes is also connected with the peak laser pulse vector potential $a_{0L}=\sum_p a^{(p)}$. To illustrate this relation, consider two lasers with identical energy. One of them is purely Gaussian with parameters matched to the blowout regime, i.e. $W_{0G}=2\sqrt{a_{0G}}$, and $P_G \propto (W_{0G} a_{0G})^2 = 4 a_{0G}^3$. Equation~(\ref{eq:energy}) then gives $\Delta E_{G}\propto a_{0G}$. In the second laser, the intensity profile is redistributed into well defined filaments sufficiently far from each other such that they cannot interact through the plasma. Consider in addition that the central laser filament spot-size is matched to the blowout radius, i.e. $W_{0L}=2\sqrt{a_{0L}}<W_{0G}$, and $P_{L}\propto (W_{0L} a_{0L})^2 = 4 a_{0L}^3$ (this condition may be satisfied after some propagation in the plasma). The corresponding energy gain is then $\Delta E_{L}\propto a_{0L}$. As a result, $\Delta E_L/\Delta E_{G} \propto a_{0L}/a_{0G}<1$. This reasoning establishes a direct relation between the maximum energy gain in LWFAs with the peak $a_0$ of lasers with higher order modes. Note however, that this is only a special case of Eq.~(\ref{eq:energy_gain_laguerre}). Still, for the case of Fig.~\ref{fig:laguerre_experimental}, $a_L = 3.18$, $a_G=3.85$, and then $\Delta E_L/\Delta E_{G} =0.82$ which is very close to simulation results, and to the prediction of the refined analytical model (generalization of Eq.~(\ref{eq:energy_gain_laguerre}) up to modes with $p=7$). In general, however, the full driver laser intensity distribution needs to be considered to predict the energy gain more acuratelly.  

Similarly to Sec.~\ref{sec:plasma}, simulations results in this section showed little correlation between the energy spread and the considered laser transverse intensity profiles. This is also due to the fact that initially the wake is not beam loaded. It is expected that in beam loaded scenarios where the energy spreads are lower, higher order modes will also affect the energy spreads. Moreover the wake excitation in the presence of higher order modes may contrast significantly with that associated with a Gaussian laser. The wake wavelength, and amplitude may then vary using non-Gaussian lasers. This may impact in the beam loading charge, in the acceleration distance, and hence in the final energy spread of the accelerated particles.

\section{Conclusions}
\label{sec:conclusions}

The influence of non-ideal plasma and laser parameters has been examined through analytical and numerical modeling. A reference simulation in a self-guided propagation regime with state-of-the-art parameters served as a baseline to address the most standard deviations to idealized parameters.

An analytical model was developed to predict the energy gain in non-homogeneous plasma density profiles. The analysis showed that the effects of the longitudinal density fluctuations are minimized when the amplitude of the fluctuations is low in comparison to the background plasma density. In addition, the density fluctuations can also be neglected when their characteristic wavelength is much shorter than the total acceleration distance. The analytical model can be generalized for arbitrary density fluctuations by writing the Fourier series of the plasma density, and performing the required integrations. However, based on simulation results using realistic plasma profiles, we expect that this will not change our conclusions. The theory also ignored the laser pulse dynamics (self-focusing, self-steepening, pump-depletion, self-compression) except for the laser pulse group velocity. Still, it captured the relevant physics and is in very good agreement with PIC simulations in QuickPIC.

The acceleration of electrons in the wake of drivers with higher order Laguerre-Gaussian modes was also investigated. An analytical model was derived to determine the electron energy gain in the wake of a higher order laser pulse. The model assumes that only the laser power inside the blowout region can effectively accelerate electrons, but neglects the details of wakefield excitation. The theory showed that higher order Laguerre-Gaussian modes generally lead to lower electron energy gains. This occurs because the laser pulse energy is typically distributed along wider radial regions in comparison to Gaussian beams. These results then emphasize that better quality focal spots lead to optimized acceleration regimes, also in agreement with previous works on self-injection~\cite{bib:ibbotson_prstab_2010}. However, the parameters of higher order Laguerre-Gaussian laser beams leading to higher energy gains were identified analytically and confirmed in QuickPIC simulations. Simulations using transverse laser profiles with larger deviations from Gaussian showed that some laser power initially at $r>W_0$ self-focused, joining the blowout region. In these situations the analytical model underestimated electron energy gain. Typically, however, the analytical and numerical modeling are in very good agreement.

It is important to remark that higher-order Gaussian beams may lead to optimized conditions for wake excitation. For instance, the ponderomotive force which is associated with the laser-pulse side filaments modifies the properties (thickness and height) of thin electron sheet which surrounds the blowout region~\cite{bib:lu_prl_2006}. Since the accelerating gradients are determined by the details of this thin sheet, the use of higher-order Laguerre-Gaussian modes may enhance the accelerating fields of the plasma wave. In addition, the use of higher-order Laguerre-Gaussian laser pulses may also lead to more effective self-guiding. For instance lasers with transverse super-Gaussian profiles may be more efficiently guided in the blowout.

\section{Acknowledgments}

The authors acknowledge fruitful discussions with Dr. N. Lopes. Work partially supported by FCT (Portugal) through grants SFRH/BD/22059/2005, PTDC/FIS/111720/2009, and CERN/FP/116388/2010, EC FP7 through LaserLab-Europe/Laptech; UC Lab Fees Research Award No. 09-LR-05-118764-DOUW, the US DOE under DE-FC02-07ER41500 and DE-FG02-92ER40727, and the NSF under NSF PHY-0904039 and PHY-0936266. Simulations were done on the IST Cluster at IST, on the Jugene supercomputer under a ECFP7 and a DEISA Award, and on Jaguar computer under an INCITE Award. 

\section*{References}
\bibliography{references}

\end{document}